\documentclass[a4paper,fleqn,usenatbib]{mnras}

\usepackage[T1]{fontenc}
\usepackage{ae,aecompl}

\usepackage{graphicx}	
\usepackage{amsmath}	
\usepackage{amssymb}	

\title[Bailey Diagrams]{Reconsidering The Bailey Diagrams of ab-type RR Lyrae Stars, Metallicity-Mediated Evolution as the Direct Cause of the Oosterhoff Phenomenon}
\author[L. -J. Li et al.]{
L. -J. Li,$^{1}$\thanks{E-mail: lipk@ynao.ac.cn}
S. -B. Qian,$^{2,3}$
L. -Y. Zhu$^{1,4}$
X. -D. Shi$^{1}$
W. -P. Liao$^{1,4}$
\\
$^1$Yunnan Observatories, Chinese Academy of Sciences, P.O.Box110, Kunming 650011, Yunnan, China \\
$^2$Department of Astronomy, School of Physics and Astronomy, Yunnan University, Kunming 650091, China \\
$^3$Key Laboratory of Astroparticle Physics of Yunnan Province, Yunnan University, Kunming 650091, People's Republic of China \\
$^4$School of Astronomy and Space Science, University of Chinese Academy of Sciences, No.1 Yanqihu East Rd, Huairou District, \\
Beijing, 101408, China\\}
\date{Accepted XXX. Received YYY; in original form ZZZ}

\pubyear{2025}

\begin{document}
\label{firstpage}
\pagerange{\pageref{firstpage}--\pageref{lastpage}}
\maketitle

\begin{abstract}

We re-examine the Bailey diagrams of fundamental mode RR Lyrae stars from the perspective of horizontal branch (HB) evolution, identifying evolutionary effects as the probable direct cause of the Oosterhoff dichotomy. By establishing empirical relationships between pulsation amplitude and average effective temperature, and utilizing pulsation period relations from nonlinear models, we transform theoretical HB evolutionary parameters into pulsation space and map them onto Bailey diagrams. We find that all pulsating Zero-Age Horizontal Branch stars fall within the Oosterhoff type I (OoI) region, with a pronounced period shift effect observed for relatively metal-rich samples ([Fe/H] $>$ -1). Evolutionary tracks confirm that OoI stars are predominantly early-stage HB stars, while Oosterhoff type II stars are highly evolved objects entering the instability strip late in their HB phase from the blue side. Crucially, metallicity plays a dual role: it directly influences pulsation periods through the period relation, but more fundamentally acts as the first parameter influencing HB morphology. This morphology statistically determines whether HB stars predominantly enter the instability strip during early or late evolutionary stages. Consequently, while evolutionary effects directly govern an individual star's position in the Bailey diagram, the population-level Oosterhoff phenomenon emerges from the interplay between these effects and the metallicity-dependent HB morphology. Our study confirms that evolutionary effects are the direct drivers of the period difference and underscores metallicity's vital role in the statistical emergence of Oosterhoff groups. Continued study of this phenomenon offers key insights into the formation history of the Milky Way and nearby dwarf galaxies.

\end{abstract}




\section{Introduction} \label{sec:intro}

RR Lyrae stars are a type of short-period pulsating variable widely distributed across various celestial structures within the Milky Way. These stars are old, low-mass objects in the horizontal branch (HB) evolutionary stage, generally believed to be over 10 billion years old \citep{2015pust.book.....C}. As a result, they are serve as important probes for studying the structure and evolution of the Milky Way and neighboring galaxies \citep{2004rrls.book.....S}. Single-mode RR Lyrae stars can be classified into two types: ab-type (RRab stars) and c-type (RRc stars). The former pulsates in the fundamental mode, while the latter pulsates in the first overtone mode, exhibiting a relatively short pulsation period. The typical lifetime of a horizontal branch star is approximately 0.1 Gyr, and it may spend only a portion of its life within the Instability Strip (IS, \citealt{1957ApJ...126..326S,1991ApJ...367..524L,2004ApJ...612..168P,2007A&A...476..307L}). For a typical HB star, evolution begins at the zero-age horizontal branch (ZAHB), progressing from blue to red, then reversing direction, and ultimately moving toward the asymptotic giant branch (AGB) located in the upper right corner of the Hertzsprung-Russel (HR) diagram. Throughout this process, the physical parameters of the stars, such as luminosity and effective temperature, change as the star evolves. Consequently, the pulsation parameters (e.g., period and amplitude), which depend on these physical parameters, will also undergo significant changes as the star evolves in the IS \citep{2004ApJ...612.1092D,2007A&A...476..307L}.

\subsection{The Oosterhoff dichotomy}

In the field of research on RR Lyrae stars, one of the significant discoveries is the Oosterhoff dichotomy. This phenomenon was identified nearly a century ago and has corresponding manifestations in various aspects, such as the difference in pulsation period observed in field stars and star clusters \citep{1939Obs....62..104O,1944BAN....10...55O,1990ApJ...350..631S,1992A&A...261..457C,2006AJ....132.1202K}, as well as the Oosterhoff gap,which reflects the evolutionary history of the Milky Way \citep{2009Ap&SS.320..261C}. Understanding this phenomenon requires consideration of both the specific evolutionary states of HB stars and the evolution of larger celestial systems, including star clusters, halos, and disks. A comprehensive study of this phenomenon also provides crucial insights into the formation and evolution of the Milky Way.

This phenomenon refers to notable differences in the average pulsation period distribution and the proportion of RRc stars across various globular clusters (GCs, \citealt{1939Obs....62..104O}). Clusters characterized by a shorter average period of RRab stars and low proportions of RRc stars (e.g., M3) are classified as Oosterhoff type I clusters (OoI clusters), while those with longer average period of RRab and higher proportions (e.g., M15) are designated as Oosterhoff type II clusters (OoII clusters, \citealt{2004rrls.book.....S}). Investigations of other Galactic GCs containing RR Lyrae stars have shown that most of these clusters belong to one of the two Oosterhoff groups. Furthermore, it has been discovered that the metallicity of stars in OoII clusters is lower than that of stars in OoI clusters \citep{1955AJ.....60..317A,1959MNRAS.119..538K}. There are also two GCs (i.e., NGC 6388 and NGC 6441) with relative high metallicity that contain long-period RRab stars, which are classified as the Oosterhoff type III clusters (OoIII clusters, \citealt{2000ApJ...530L..41P,2009Ap&SS.320..261C}). The Oosterhoff dichotomy exhibits itself on multiple scales. The $\langle P_{\rm ab} \rangle$ - [Fe/H] diagram for GCs in the Milky Way reveals a distinct separation between OoI and OoII GCs, and the region between them is referred to as the Oosterhoff gap \citep{1990ApJ...350..631S,2004ASPC..310..113C,2009Ap&SS.320..261C}. However, RRab stars in the dwarf satellite galaxies of the Milky Way and their associated GCs tend to occupy the Oosterhoff gap region, creating a stark contrast with the GCs found in the Milky Way \citep{2009Ap&SS.320..261C}. The Oosterhoff gaps illustrated in the $\langle P_{\rm ab} \rangle$ - [Fe/H] and HBt - [Fe/H] diagrams are more closely related to the formation and evolution of the Milky Way's substructure \citep{2009Ap&SS.320..261C,2024MNRAS.534.3654P}.

Mounting evidence from contemporary astronomical surveys and observations indicates that there are indications of a connection between the Oosterhoff dichotomy and merging events in the Milky Way \citep{2019ApJ...882..169F,2021ApJ...919..118F,2021MNRAS.502.5686I,2022MNRAS.517.2787L,2023MNRAS.525.5915Z,2024A&A...690L..17L}. A deeper understanding of the Oosterhoff phenomenon may enhance our study of the formation and evolution of galactic substructures \citep{2015ApJ...798L..12F,2015ApJ...811..113P,2018MNRAS.477.1472B,2019MNRAS.484.4833P,2020A&A...635A..66P,2020AcA....70..121P,2020MNRAS.492.2161Z,2021MNRAS.502.5686I,2022MNRAS.517.2787L,2022MNRAS.513.1958W}.

\subsection{Mechanism}
Numerous studies have been conducted to elucidate the Oosterhoff dichotomy. The corresponding mechanisms include the hysteresis zone, metallicity, helium enhancement, and the evolutionary effects on which this article focuses. We will introduce each of these mechanisms separately.

\citet{1973ApJ...185..477V} proposed the existence of a hysteresis zone in the RR Lyrae IS and utilized this concept to explain the proportions of RRab and RRc stars in different Oosterhoff GCs. Within the hysteresis zone, the pulsation mode depends upon the direction of stellar evolution: blueward evolution predominantly produces RRab stars, as observed in OoI clusters; while redward evolution produces more RRc stars, as seen in OoII clusters. In subsequent theoretical research, this zone was referred to as the "OR region", its definition has gradually evolved. It is now understood as the region where both fundamental and first overtone modes exhibit a stable limit cycle \citep{1997A&AS..121..327B,2015ApJ...808...50M}, with increasing attention also directed toward double-mode RR Lyrae stars (RRd stars). However, due to both observational and theoretical uncertainties existing, observations have neither confirmed nor disproven the existence of the zone as originally defined \citep{2004rrls.book.....S}.

\citet{1958RA......5...41S} suggested that OoII RR Lyrae stars are brighter than those in OoI GCs. Subsequently, \citet{1981ApJS...46...41S} discovered the period shift effect based on observations of RR Lyrae stars in GCs M3 and M15. In later studies, \citet{1981ApJ...248..161S} and \citet{1982ApJ...252..553S} pointed out that this effect is present in other GCs and exhibits a general correlation with metallicity. The period shift is considered to exist for each RR Lyrae stars in a given cluster, and field RR Lyrae stars also demonstrate the period shift-metallicity effect \citep{1990ApJ...350..631S}. Sandage's explanation of the Oosterhoff phenomenon and the period shift effect primarily relied on an increase in RR Lyrae luminosity with decreasing metallicity \citep{2004rrls.book.....S}. However, this perspective assumes an undesirable anticorrelation between helium abundance $Y$ and metallicity $Z$ \citep{1990ApJ...350..631S,1992A&A...261..457C}. Recently, with the acquisition of the spectral metallicity data for RR Lyrae stars (see series of papers of \citealt{2019ApJ...882..169F} and \citealt{2020ApJS..247...68L}), the correlation between metallicity and pulsation period has been further confirmed, although the intrinsic relationship remains poorly understood.

In recent years, multiple populations and helium enhancements have been employed to explain this phenomenon in GCs (see \citealt{2014MNRAS.443L..15J}, \citealt{2021ASPC..529..169L} and references therein). \citet{2014MNRAS.443L..15J} pointed out that the metal-rich OoI clusters are primarily formed by first-generation stars without enhancements, while the metal-poor OoII clusters are populated by second-generation stars with enhanced helium and CNO abundances.

The evolutionary effect has been recognized in previous investigations as well. For instance, Sandage has also mentioned the impact of evolution on the variability of starlight in star clusters \citep{1981ApJ...244L..23S}. In the hysteresis zone theory proposed by \citet{1973ApJ...185..477V}, the various evolutionary paths are crucial for supporting their perspective. A significant contribution that explicitly links the Oosterhoff phenomenon to evolutionary effects is the work of \citet{1990ApJ...350..155L}. Based on their calculations, they stated that "almost all of the RR Lyrae variables in group II clusters are highly evolved stars from the blue side of the instability strip" and that "calculations for group I clusters like M3 indicate that the masses and luminosities of the RR Lyrae variables are not much different from those of the ZAHB models". The notion that evolution is related to the Oosterhoff dichotomy is further supported by both models and observations \citep{1986A&A...169..111G,1992AJ....104..253C,1995ASPC...78..243C,1999ApJ...515L..85C,1999AJ....118.1373L,2005AJ....129..267C,2020A&A...635A..66P,2023MNRAS.525.5915Z}.
However, there remains no unified consensus regarding the mechanisms underlying this phenomenon \citep{2004rrls.book.....S,2015pust.book.....C,2021ApJ...919..118F}. Even in recent research, scholars maintain differing opinions on the existence of the Oosterhoff dichotomy \citep{2024A&A...684A.173C,2024MNRAS.534.3654P}.

\subsection{Present Paper}

In this paper, we focus on relevant work from the evolutionary perspective. The key issue of the Oosterhoff dichotomy is the difference in pulsation period, which, according to the pulsation model \citep{1971ApJ...169..311V,1997A&AS..121..327B,2015ApJ...808...50M}, is closely related to the evolutionary parameters such as mass, effective temperature, and luminosity. Therefore, understanding the mechanism behind the Oosterhoff phenomenon requires a thorough examination of the evolution of RR Lyrae stars. While the direct impact of metallicity on the pulsation period is limited, it still plays an important role in the Oosterhoff phenomenon. As the first parameter, metallicity influences the HB morphology leading to variations in the distribution of physical parameters among HB stars with different metallicities. When combined with evolutionary effects, the phenomenological relationship between the observed pulsation period and metallicity ultimately emerges \citep{2019ApJ...882..169F,2021ApJ...919..118F}.

Some scholars have sought to validate the theory of evolutionary effects from an observational standpoint by plotting different Oosterhoff type RR Lyrae stars on the HR diagram and comparing their positions (e.g., \citealt{2019MNRAS.484.4833P}). This process requires the acquisition of their temperatures, luminosities, color indices, and apparent magnitudes. However, due to the inherent pulsations of these stars, these parameters can fluctuate, which limits the accuracy of the measurements obtained. In contrast, the accuracy of pulsation parameters is generally higher, as they are independent of distance and reddening effects. Therefore, by establishing the relationships between evolutionary parameters and pulsation parameters, we can delineate the theoretical evolutionary tracks of HB stars on the Bailey diagram (i.e., Period-Amplitude diagrams) and intuitively compare them with observational results. In this context, the Bailey diagram can be regarded as an HR diagram with pulsation parameters as the variables. It is commonly employed to investigate the Oosterhoff phenomenon present in Galactic field stars and other substructures \citep{2006AJ....132.1202K,2009AJ....138.1284K,2009AcA....59..137S,2011rrls.conf...17S,2019ApJ...882..169F,2020ApJ...896L..15B,2021ApJ...919..118F,2023MNRAS.525.5915Z,2024A&A...690L..17L}.

In Section \ref{sec:relation}, we utilize the empirical relations provided to establish the relationship between physical evolution parameters and pulsation parameters, especially between effective temperature ($T_{\rm eff}$) and the amplitude of the G band ($A$(G)). The period relations are derived using pulsation models (see \citealt{2015ApJ...808...50M} and references therein). The amplitude relation is based on theoretical perspectives and empirical equations provided by \citet{1981ApJ...244L..23S,1997A&AS..121..327B,2014AJ....147...31M}. In Section \ref{Sec:BaileyDiagram}, we utilize the relationships between pulsation parameters and physical parameters to derive the pulsation characteristics of RR Lyrae stars across various horizontal branching states. Subsequently, we illustrate the ZAHB lines and evolutionary tracks on the Bailey diagram. Sections \ref{Sec:Discussion} and \ref{Sec:Conclusion} present the discussion and summary, respectively.

\section{Relationship between $T_{\rm eff}$ and $A$(G)} \label{sec:relation}

Given the structural evolution parameters (mass, [Fe/H], $T_{\rm eff}$, luminosity, etc.) of a horizontal branch star in the IS, its pulsation period can be obtained by using the classic equation provided by \citet{1971ApJ...169..311V}. However, this relation was established 50 years ago based on linear non-adiabatic pulsation models. Consequently, we adopt a more recent relation provided by \citet{2015ApJ...808...50M}:

\begin{equation}
 \begin{aligned}
\log P_{\rm pul} &= 11.347 + 0.860 \log \frac{L}{L_{\rm \odot}} - 0.58 \log \frac{M}{M_{\rm \odot}} \\
 &- 3.43 \log T_{\rm eff} + 0.024 \log Z.\label{Equ:1}
 \end{aligned}
\end{equation}

It is derived based on nonlinear pulsation models that encompass a broad range of metal abundances (Z = 0.0001 - 0.002). Our next step is to investigate the correlation between pulsation amplitude and these evolutionary parameters. \citet{1981ApJ...244L..23S} pointed out that the $B$ band amplitude, $A$($B$), is not a function of luminosity or [Fe/H], but primarily depends on $T_{\rm eff}$ alone. \citet{1997A&AS..121..327B} employed a nonlinear convection model to conduct theoretical research, which revealed that luminosity amplitude is closely related to $T_{\rm eff}$, while the influences of stellar mass and metal abundance can be disregarded. \citet{2014AJ....147...31M} also emphasized that the dependence of the light amplitude of RR Lyrae stars on temperature is independent of Oosterhoff type. They provided equations that describe the relationships between the intrinsic color index $(B-V)_0$ and the amplitudes of the $B$ and $V$ bands, which can be utilized to determine interstellar reddening. Based on these literature, we assume that the average effective temperature is the sole variable in the amplitude function and aim to establish the relationship between these two parameters.

We employed two approaches to establish this relationship. The first method utilizes observational data, while the second relies on empirical relations. For the former, the $T_{\rm eff}$ data are sourced from \citet{2019ApJS..245...34X}, and the amplitude data we used are the G band amplitude $A$(G) provided by Gaia DR3 \citep{2022yCat.1358....0G,2023A&A...674A..18C}. However, during our study, we recognized that various factors, such as the time resolution of spectral observations, pulsation phase, and pulsation modulations, can influence both $T_{\rm eff}$ and $A$(G), complicating the analysis. Consequently, we ultimately opted for empirical equations to derive the relation between $T_{\rm eff}$ and $A$(G), which are:
\begin{equation}
A(\rm G) = 0.925 \emph{A}_{\emph{V}} - 0.012,\label{Equ:2}
\end{equation}
\begin{equation}
(\langle B \rangle - \langle V \rangle)_{0} = -0.078 A^{2}_{V} - 0.014 A_{V} + 0.395,\label{Equ:3}
\end{equation}
and
\begin{equation}
\log T_{\rm eff} = -0.309 (B - V)_{0} + 3.919.\label{Equ:4}
\end{equation}
Equation \ref{Equ:2} is derived from Equ. (3) of \citet{2016A&A...595A.133C}, while Equation \ref{Equ:3} is based on Equ. (5) of \citet{2014AJ....147...31M}. Equation \ref{Equ:4} is also derived from Equ. (3) of \citet{2014AJ....147...31M} and is based on the color and temperature data provided by \citet{1999A&A...346..564C} and \citet{2005AJ....129..267C}. In Equations \ref{Equ:2} and \ref{Equ:3}, $A_{V}$ represents the pulsation amplitude in the V-band. According to the aforementioned relations, when the minimum $A(\rm G)$ is 0.1 mag and the maximum $A(\rm G)$ is 1.3 mag, the corresponding $\log T_{\rm eff}$ are 3.798 and 3.852, respectively. This indicates that the width of the fundamental mode IS is approximately 830 K.

\begin{figure}
\includegraphics[width=0.45\textwidth]{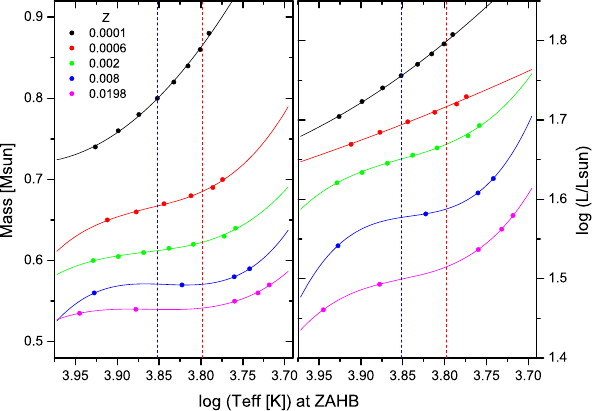}
\caption{Synthetic ZAHB of different [Fe/H] near the RRab IS (solid points). The solid lines represent the corresponding polynomial fitting curves. For better visualization, only five metal abundance scenarios are depicted. The BaSTI $\alpha$-enhanced canonical model is utilized \citep{2006ApJ...642..797P}.} \label{Fig.1}
\end{figure}

\section{Bailey Diagram} \label{Sec:BaileyDiagram}

After establishing the relationship between pulsation parameters and evolutionary parameters, the focus now shifts to obtaining the evolutionary parameters of HB stars. In this paper, we utilize the theoretical model parameters provided by the Bag of Stellar Tracks and Isochrones (BaSTI) $\alpha$-enhanced canonical model\footnote{\url{http://albione.oa-teramo.inaf.it/}} \citep{2006ApJ...642..797P}.

\citet{2006ApJ...642..797P} provided an extensive set of stellar evolution models and isochrones for an $\alpha$-enhanced metal distribution, which is consistent with observations of the Galactic halo and bulge populations. They also presented comprehensive sets of evolutionary models for low-mass, He-burning stellar structures, covering a metallicity range from $Z$ = 0.00001 to 0.05, including the ZAHB sequences and evolutionary sequences for dozens of masses for each composition\footnote{In this paper we used the models with mass loss efficiency $\eta$ = 0.4.}. Each stellar model parameters include luminosity, $T_{\rm eff}$, $M_{V}$ and color indices. Our focus is on parameters located in and near the IS; however, the data available in the ZAHB sequences and the central helium exhaustion loci provided by \citet{2006ApJ...642..797P} in this region are limited. To quantify the relationship between mass (luminosity) and $T_{\rm eff}$, we assume that the ZAHB and the central helium exhaustion lines are continuous and smooth, and utilize the evolutionary parameters (mass, luminosity, and $T_{\rm eff}$) within and near the IS as fitting objects. We employ linear polynomials as fitting formulas to obtain the mass - $T_{\rm eff}$ and luminosity - $T_{\rm eff}$ relationships for different metallicities in this region. Figure \ref{Fig.1} illustrates the theoretical ZAHB in and near the IS. As shown in the figure, the fitted curve closely aligns with the data points, indicating that the polynomial model effectively captures the overall trend of parameter variation, with a coefficient of determination (R$^{2}$) larger than 0.9. These relationships are utilized to determine the mass and luminosity, which subsequently allow us to calculate the $P_{\rm pul}$. In this manner, we can plot the continuous ZAHB lines and central helium exhaustion loci in the
Bailey diagram. The BaSTI HB evolutionary tracks offer a greater variety of stellar models, ensuring a sufficient number of points in the IS. We can directly utilize these data to plot the evolutionary tracks in the Bailey diagram by applying the relations mentioned in Section \ref{sec:relation}. Additionally, BaSTI provide the corresponding horizontal branch ages of stars, which can be used to evaluate the duration that HB stars can remain in the IS.

\begin{table}
\caption{Evolutionary parameters of ZAHB RRab stars at $A$(G) = 0.70 mag (i.e., $\log T_{\rm eff}$ = 3.814).}\label{Table1}
\begin{tabular}{l c c c c}
\hline\hline
$Z$	     & [Fe/H] & $\log (L/L_{\odot})$ &  Mass  &  $\log P_{\rm pul}$ \\
         &   dex  &      dex            &  $M_{\rm \odot}$ &  dex\\
\hline
0.0001 & -2.62  &  1.785 &  0.843 & -0.2544   \\
0.0003 & -2.14  &  1.727 &  0.721 & -0.2534   \\
0.0006 & -1.84  &  1.710 &  0.678 & -0.2458   \\
0.001  & -1.62  &  1.688 &  0.648 & -0.2480   \\
0.002	 & -1.31  &  1.663 &  0.618 & -0.2503   \\
0.004	 & -1.01  &  1.628 &  0.593 & -0.2628   \\
0.008	 & -0.70  &  1.583 &  0.570 & -0.2836   \\
0.010	 & -0.60  &  1.561 &  0.562 & -0.2968   \\
0.0198 & -0.29  &  1.509 &  0.540 & -0.3246   \\
\hline\hline
\end{tabular}
\end{table}

\begin{figure*}
\centering
\includegraphics[width=0.65\textwidth]{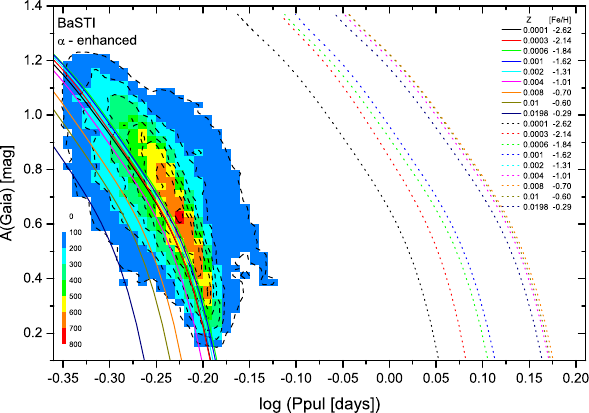}
\caption{Density maps of the Bailey diagram for Gaia DR3 RRab stars. The solid lines on the left show the ZAHB for different [Fe/H]. The dashed lines diapaly isodensity levels in increments of 100. Various blocks represent distinct numerical densities. The dotted lines represent the theoretical predictions for central Helium exhaustion. The BaSTI $\alpha$-enhanced canonical model is utilized \citep{2006ApJ...642..797P}. Some curves overlap with each other, and readers can zoom in on the image to examine the details.} \label{Fig.2}
\end{figure*}

\subsection{ZAHB in Bailey Diagram} \label{Sec:BaileyDiagramSub1}

In Figure \ref{Fig.2}, the solid lines represent the ZAHB for different [Fe/H] on the Bailey diagram. For comparison, we also present the color density map of the Gaia DR3 RRab sample. The contour lines, indicated by dashed lines, illustrate isodensity levels in increments of 100. Different colored blocks signify distinct numerical densities. The density maps in Figure \ref{Fig.2} primarily reflect the distribution of OoI-type stars, which constitute the majority of RRab stars, while most ZAHB lines align with the distribution characteristics of OoI stars. However, there is a slight deviation between the solid lines and the region of highest concentration. This can be understood in a manner similar to how most main sequence stars are positioned above the zero-age main sequence in the HR diagram; most OoI stars are in the "main sequence stage" of the HB. They exhibit a slightly higher degree of evolution than the ZAHB, and are consequently located to the right of the ZAHB lines in the Bailey diagram.

\begin{figure}
\includegraphics[width=0.45\textwidth]{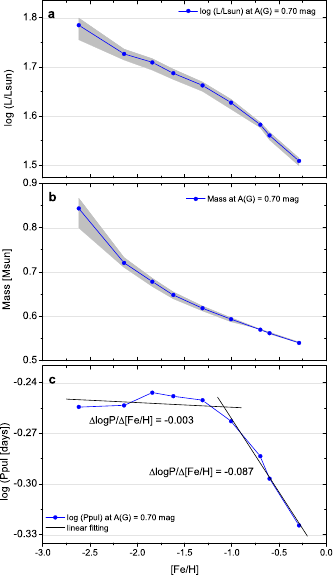}
\caption{Dependencies between evolutionary parameters and metallicity of ZAHB RRab stars with a amplitude of 0.70 mag (i.e., $\log T_{\rm eff}$ = 3.814). Panel (a): $\log L/L_{\rm \odot}$ vs. [Fe/H]; Panel (b): Mass vs. [Fe/H]; Panel (c): $\log P_{\rm pul}$ vs. [Fe/H]. The shaded areas in panel (a) and (b) denote the luminosity and mass range of ZAHB RRab stars. The solid lines in panel (c) represent a segmented linear fitting of the data.} \label{Fig.3}
\end{figure}

In Figure \ref{Fig.2}, it can be observed that when [Fe/H] $<$ -1, the distributions of the solid lines corresponding to different [Fe/H] values are not significantly different; they overlap and are even difficult to distinguish. However, when [Fe/H] $>$ -1, as [Fe/H] increases, the lines shift towards the left side of the diagram, exhibiting the period shifting phenomenon. We present the luminosities, masses, and corresponding $\log P_{\rm pul}$ of ZAHB RRab stars with a pulsation amplitude of 0.70 mag (i.e., $\log T_{\rm eff}$ = 3.814) at different metellicities in Table \ref{Table1}, and illustrate the relationships between these parameters and [Fe/H] in Figure \ref{Fig.3}. As shown in panels (a) and (b) of Figure \ref{Fig.3}, there are significant differences in the masses and luminosities of pulsating ZAHB stars with varying [Fe/H], with both parameters decreasing as [Fe/H] increases. In panel (c) of Figure \ref{Fig.3} (i.e., the $\log P_{\rm pul}$ - [Fe/H] diagram), the observed relationship further confirms the distribution of ZAHB lines depicted in Figure \ref{Fig.2}. The $\log P_{\rm pul}$ and [Fe/H] are not simply linearly related; rather, there is a turning point near [Fe/H] = -1. Segmented linear fitting indicates that when [Fe/H] $<$ -1, $\Delta \log P_{\rm pul}/\Delta$[Fe/H] = -0.003(5), and when [Fe/H] $>$ -1, $\Delta \log P_{\rm pul}/\Delta$[Fe/H] = -0.087(7) (see the solid lines in panel (c) of Figure \ref{Fig.3}). The former result is consistent with \citet{1987ApJ...312..762S}, while the latter suggests a significant periodic shift phenomenon. However, the theoretical period shifting occur in the OoI region and cannot explain the observed period shifting \citep{1990ApJ...350..631S}; thus, and other factors (such as longer pulsation period caused by evolution) must be considered.

The reason for this phenomenon can be attributed to the pulsation relation. For a constant amplitude (i.e., constant $T_{\rm eff}$), Equation (\ref{Equ:1}) provided by \citet{2015ApJ...808...50M} implies that
\begin{equation}
\Delta \log P_{\rm pul} = 0.86 \Delta \log \frac{L}{L_{\rm \odot}} - 0.58 \Delta \log \frac{M}{M_{\rm \odot}} +0.024 \Delta \log Z.\label{Equ:5}
\end{equation}
From this, it can be observed that the pulsation period is not only proportional to the logarithm of luminosity but also inversely proportional to the logarithm of mass. Therefore, for stars with [Fe/H] $<$ -1.0, as [Fe/H] increases, both their luminosities and masses decrease (see Figure \ref{Fig.3}). The effects of these three parameters offset each other, resulting in a minimal impact on the pulsation periods. In contrast, for stars with [Fe/H] $>$ -1.0, the luminosity factor dominates, leading to significant variations in $P_{\rm pul}$. The most metal-rich ZAHB line is located at the far left of the Bailey diagram, which significantly deviates from the density map distribution.

The dotted lines on the right side of Figure \ref{Fig.2} represent the theoretical predictions for central helium exhaustion. They approximately illustrate the upper bounds for the periods that these stellar structures can exhibit throughout their entire HB evolution. It can be observed that the upper limit of the pulsation period is greater for metal-rich stars. This indicates that, compared to metal-poor RRab stars, metal-rich stars are more widely distributed in the Bailey diagram (refer to Figure 35 in \citealt{2023A&A...674A..18C}).

\begin{figure*}
\centering
\includegraphics[width=0.9\textwidth]{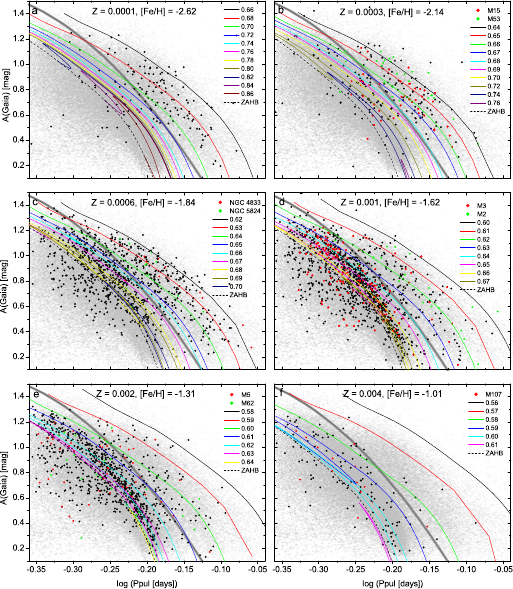}
\caption{Bailey diagrams of different chemical compositions. In each panel, the gray background dots represent Gaia DR3 RRab stars, the black solid dots represent field RRab stars with corresponding [Fe/H] \citep{2020ApJS..247...68L}, and the red and green solid dots represent RRab stars in GCs \citep{2001AJ....122.2587C,2017yCat.5150....0C}. The black dashed line in each panel refers to the ZAHB of the corresponding [Fe/H]. The solid lines represent the evolutionary tracks of stars with different masses. The thick gray solid lines present the division of OoI and OoII regions \citep{2021ApJ...919..118F}.} \label{Fig.4}
\end{figure*}

\begin{figure*}
\centering
\includegraphics[width=0.9\textwidth]{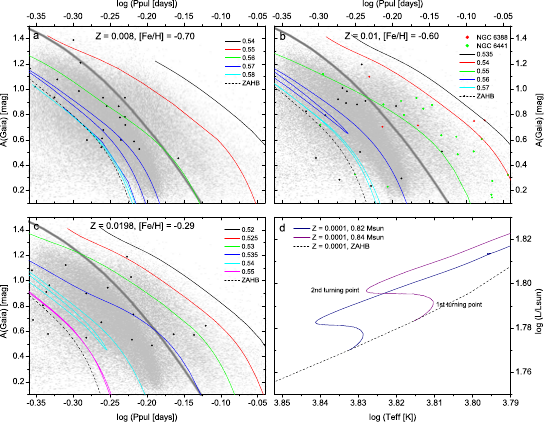}
\caption{Panel (a)-(c): Same as Figure \ref{Fig.4}, but Bailey diagrams for relatively metal-rich ([Fe/H] $>$ -1.0). Panel (d): Evolutionary tracks for Z = 0.0001, [Fe/H] = -2.62. The two turning points are labeled.} \label{Fig.5}
\end{figure*}

\subsection{Evolutionary tracks in Bailey Diagram} \label{Sec:BaileyDiagramSub2}

We present the evolutionary tracks of HB stars with varying metallicities in the Bailey diagrams (see Figures \ref{Fig.4} and \ref{Fig.5}). The black dashed lines in each panel represent the ZAHB of the corresponding [Fe/H]. The solid lines, depicted in different colors, illustrate the evolution of stars with varying masses. The thick gray solid lines delineate the boundaries between the OoI and OoII regions \citep{2021ApJ...919..118F}, while the black solid dots represent RRab stars with [Fe/H] values similar to those listed in the individual panels (data provided by \citealt{2020ApJS..247...68L}). The red and green solid dots indicate RRab stars located in GCs \citep{2001AJ....122.2587C,2017yCat.5150....0C}. These stars are plotted in the corresponding panels based on the [Fe/H] of their host clusters. Specifically, the samples originate from M15, M53, NGC 4833, NGC 5824, M2, M3, M62, M5, M107, NGC 6388, and NGC 6441. The corresponding mean [Fe/H] values are as follow: [Fe/H] = -2.26, -1.99, -1.88, -1.85, -1.62, -1.57, -1.29, -1.27, -1.04, -0.60, and -0.53 \citep{2009Ap&SS.320..261C}. The gray background dots represent RRab samples sourced from Gaia DR3 \citep{2022yCat.1358....0G,2023A&A...674A..18C}.

Figures \ref{Fig.4} and \ref{Fig.5} provide extensive information and details. In the OoI region, the evolutionary tracks exhibit considerable complexity. In the HR diagram, the HB stars initially move slightly from ZAHB toward the red region, then shift toward the blue (the first turning point), and ultimately progress toward the AGB with increased luminosity (the second turning point, see Panel (d) of Figure \ref{Fig.5}). The lines representing masses of 0.82, 0.84, and 0.86 $M_{\odot}$ in Panel (a) of Figure \ref{Fig.4} illustrate the corresponding evolutionary tracks. Specifically, in Panel (b) of Figure \ref{Fig.4}, the dark yellow upward curve (0.72 $M_{\odot}$) indicates the first turning point, while the right dark yellow line represents the track following the second turning point. The navy and purple downward curves (0.74 and 0.76 $M_{\odot}$) signify the second turning point. The corresponding stars are relatively massive, with temperatures at the ZAHB stage lower than the red boundary of the IS. These stars will enter and subsequently exit the IS during their evolutionary processes. Additionally, the evolutionary lines between the two turning points are depicted in the panels (the dark yellow line on the left in Panel (c), the yellow line on the left in Panel (d), and the cyan lines on the left in Panels (e) and (f)). These lines evolve along the ZAHB toward the upper left, consistent with the distribution of OoI stars. Evolutionary tracks on the right side of each panel represent the late stage of the HB, specifically evolving toward AGBs, with the evolutionary direction progressing from the top left to the bottom right. Their positions align with those of the stars in the OoII group. The masses of the corresponding stars are relatively smaller than those of RRab stars with the same metallicity in the OoI region.

\begin{figure}
\includegraphics[width=0.45\textwidth]{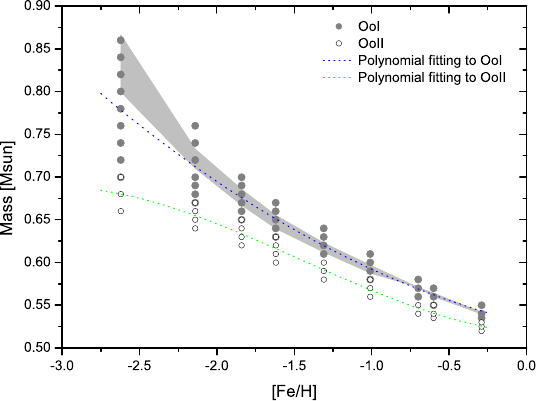}
\caption{The masses of OoI and OoII RRab stars versus the [Fe/H]. The dotted lines represent the corresponding polynomial fitting curves. The shaded area denotes the mass range of ZAHB RRab stars.} \label{Fig.6}
\end{figure}

\begin{figure}
\includegraphics[width=0.45\textwidth]{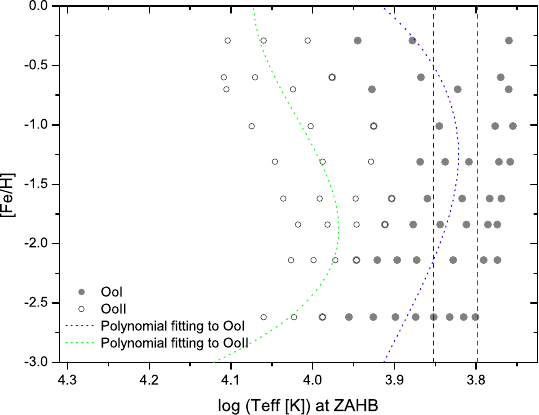}
\caption{The $T_{\rm eff}$ of OoI and OoII RRab stars at ZAHB versus the [Fe/H]. The dotted lines represent the corresponding polynomial fitting curves, and the dashed vertical lines represent the boundaries of RRab IS.} \label{Fig.7}
\end{figure}

In Figure \ref{Fig.6}, we plotted the masses of OoI and OoII stars against varying [Fe/H] values and provided corresponding polynomial fitting curves. This analysis indicates a significant negative correlation between mass and [Fe/H]. This relationship elucidates why, overall, metal-poor OoII stars typically exhibit a higher mass than metal-rich OoI stars \citep{2019MNRAS.484.4833P}. However, it is crucial to note that when metallicity is held constant, the mass of OoI stars is actually greater than that of OoII stars with the same metallicity. This demonstrates that higher mass is not an intrinsic characteristic of OoII stars, but rather a surface phenomenon associated with their generally low metallicity. As illustrated in Figure \ref{Fig.6}, the mass of metal-poor OoII stars may be comparable to or even greater than that of relatively metal-rich OoI stars.

The temperatures of OoII stars remained above the blue boundary of the IS during the early stages of HB evolution, only entering the IS from the left side with higher luminosities in the later stages. Figure \ref{Fig.7} displays the $T_{\rm eff}$ of OoI and OoII stars at their ZAHB, where the blue and green dotted lines represent the corresponding polynomial fitting curves. It is evident that the ZAHB $T_{\rm eff}$ of OoII stars exceed 8000 K ($\log T_{\rm eff}$ $\simeq$ 3.90), indicating that they were A-type HB stars (HBA stars, \citealt{2009Ap&SS.320..261C}). Furthermore, the green dotted line illustrates that the ZAHB $T_{\rm eff}$ also vary with [Fe/H]: in the cases of the most metal-rich and metal-poor stars, the $T_{\rm eff}$ of OoII stars at the ZAHB stage are systematically higher than those of stars with intermediate metallicities.

Figure \ref{Fig.5} illustrates the evolutionary tracks of RRab stars with relatively high metallicity. The ZAHB loci of these stars are positioned on the left side of the diagram, resulting in their evolutionary tracks being located in both the OoI and OoII regions during the later stages of evolution. This suggests that these metal-rich RRab stars are widely distributed across the Bailey diagram. The study of \citet{2021ApJ...919..118F} includes a limited number of RRab samples with [Fe/H] $>$ -1.0, represented by warm-colored points in their Figure 7. It is evident that their distribution on the Bailey diagram is more extensive compared to that of other metal-poor samples. Furthermore, a notable decline in the sample size is observed in Figure \ref{Fig.5}; however, this reduction is likely attributable to observational effects and data processing strategies. Most of the data in \citet{2020ApJS..247...68L} were sourced from the Large Sky Area Multi-Object Fiber Spectroscopic Telescope (LAMOST) survey, which primarily targets halo stars rather than those in the Galactic disk and bulge. Additionally, the matching template employed in their data processing imposes constraints on [Fe/H]. It can be inferred that there are likely more relatively metal-rich RR Lyrae stars in the disk \citep{2020MNRAS.492.2161Z}, which may have formed through atypical channels \citep{2024MNRAS.52712196B}.

The observed results indicate a significant difference in the [Fe/H] values between OoI and OoII stars (see \citealt{2019ApJ...882..169F,2021ApJ...919..118F,2023MNRAS.525.5915Z}, and references therein). However, each panel of Figure \ref{Fig.4} displays both OoI and OoII types of samples exist simultaneously, albeit in varying proportions. This indicates that the relationship between [Fe/H] and $P_{\rm pul}$ is a phenomenological statistical result of the population. According to the pulsation model, physical evolution parameters such as mass, $T_{\rm eff}$, luminosity, and [Fe/H] collectively determine the value of $P_{\rm pul}$, with [Fe/H] being just one of these factors. In this context, the direct mechanism underlying the Oosterhoff period dichotomy appears to be an evolutionary effect. However, the important role of [Fe/H] is manifested in other aspects, as it shapes the HB morphology of GCs as the first parameter. This influence results in the statistical Oosterhoff phenomenon at the population level, which includes differences in the average periods of GC RRab stars, the proportion of GC RRc stars, and the issues related to the OoIII group. For a detailed discussion, please refer to Section \ref{Sec:Discussion}.

In Figures \ref{Fig.4} and \ref{Fig.5}, it can be observed that some stars lie below the ZAHB lines. Several possible explanations account for this phenomenon: 1) measurement errors in [Fe/H], which may suggest that the [Fe/H] of the stars is actually higher, resulting in a leftward shift in the diagram; 2) the pulsations of the stars may exhibit Blazhko modulations \citep{1907AN....175..325B}, which are generally believed to suppress pulsation, leading to a measured amplitude that is smaller than that of non-Blazhko stars with the same pulsation period (refer to Page 106 of \citealt{2004rrls.book.....S}); and 3) the presence of companions whose luminosity cannot be disregarded, as the additional light causes the observed amplitude to decrease \citep{2020RAA....20...94L}.

\section{Discussions}  \label{Sec:Discussion}

\subsection{Comparison with other Bailey diagrams} \label{Sec:Discussion:part1}
The practices of plotting evolutionary tracks in Bailey diagrams have already been documented in \citet{1997A&AS..121..327B} and \citet{2020ApJ...896L..15B}. The track shapes in Figures 16 and 17 of \citet{1997A&AS..121..327B} are generally consistent with our results presented in Panels (a) and (b) of Figure \ref{Fig.4}. The primary difference is that the ZAHB line in their Figure 17 is positioned further to the right compared to ours. \citet{2020ApJ...896L..15B} provided several examples of stellar evolution, specifying only one mass condition specified for each metallicity. The evolutionary tracks in their Bailey diagrams align with those in the present paper. However, the mass distribution of RR Lyrae stars in GCs should follow a specific pattern, and relying on a single evolutionary track is inadequate. Similar studies have been conducted by \citet{2010ApJ...722...79S}, but they adopted the same approach, where each GC is represented by only one evolutionary track. Our advantage is that we provide the Bailey diagram of pulsating ZAHB stars with different metallicity (see Figure \ref{Fig.2}), as well as the evolutionary tracks of different masses (see Figures \ref{Fig.4} and \ref{Fig.5}). As introduced in Section \ref{Sec:BaileyDiagram}, our figures sufficiently describe the evolution of RRab stars in the Bailey diagram. The theoretical tracks align well with observational data and can elucidate some phenomena that were previously challenging to explain. By following this analytical approach, future researchers can develop more detailed theoretical models, construct more accurate evolutionary trajectories, and conduct more in-depth  investigations into specific aspects of the Bailey diagram, such as the HASP \citep{2014PASP..126..616S,2015ApJ...798L..12F,2018MNRAS.477.1472B}, and the different populations \citep{2015ApJ...811..113P,2020AcA....70..121P}.

\subsection{Important roles of metallicity and HB morphology} \label{Sec:Discussion:part2}
For an RR Lyrae star, its position in the Bailey diagram is directly influenced by evolutionary factors. However, in celestial systems composed of numerous stars, such as GCs, substructures of the Milky Way, and nearby dwarf galaxies, the appearance of their Bailey diagrams is also affected by additional factors, among which the most important ones are metallicity and the HB morphology it affects. The HB structure parameter (HBt) describes the HB morphology, expressed as $(B-R)/(B+V+R)$ \citep{1990ApJ...350..155L,1994ApJ...423..248L}, where $B$ represents the number of HB stars to the blue of the IS, $V$ denotes the number of RR Lyrae variables, and $R$ indicates the number of HB stars to the red of the IS. \citet{2024MNRAS.534.3654P} collected HBt values of multiple GCs and plotted the HBt-[Fe/H] diagram. In that diagram, almost all GCs with low metallicity ([Fe/H] $<$ -1.50) exhibit high HBt values ($>$ 0.5) and belong to the OoII group (see their Figure 4). This finding suggests that metallicity is the first parameter determines the HB morphology of a GC \citep{1960ApJ...131..598S,2009Ap&SS.320..261C}. Furthermore, a high HBt value indicates that the corresponding GC has a significant number of blue HB components. As these member stars evolve, they will eventually enter the IS from the blue side and become the OoII RRab stars with long periods, which constitute the majority of variable stars. Additionally, the $T_{\rm eff}$ of RRc stars are higher than those of RRab stars and are closer to the blue HB region. Consequently, the proportion $N_{\rm RRc}/N_{\rm tot}$ is greater than that of GCs with redder HB morphology (also refer to Section 7 of \citealt{2021ApJ...919..118F}). Therefore, the interplay between evolutionary effects and HB morphology can effectively explain the Oosterhoff groups (i.e., differences in average period of RRab stars and population ratio).

Halo RR Lyrae stars, originating either from distinct accretion events or as in-situ populations, exhibit distinct morphological features in their corresponding Bailey diagrams \citep{2024A&A...690L..17L}. This difference likely arises from variations in the metallicity distribution function (MDF) of these populations, particularly the dispersion of their [Fe/H] distribution. For instance, stars associated with the largest known accretion event, the Gaia-Sausage-Enceladus (GS/E, \citealt{2018MNRAS.478..611B,2018Natur.563...85H}), display a narrow MDF centred at [Fe/H] = -1.17 with a dispersion of 0.34 dex \citep{2020MNRAS.497..109F}. In contrast, the halo RR Lyrae population as a whole (comprising both accreted and in-situ stars), typically exhibits a relatively lower mean metallicity and spans a broad metallicity range \citep{2019ApJ...882..169F,2020ApJS..247...68L,2021ApJ...919..118F,2022MNRAS.517.2787L,2023MNRAS.525.5915Z}. A larger metallicity dispersion corresponds to a prolonged star formation and chemical evolution history, resulting in a broader distribution of stellar physical parameters. This, in turn, leads to a greater dispersion in pulsation period. Conversely, systems characterised by narrow metallicity dispersions (e.g., GS/E and halo GCs) exhibit more tightly clustered physical and evolutionary parameters, manifested as clearly delineated sequences in the Bailey diagram.

\subsection{Metal-rich OoIII group GCs}  \label{Sec:Discussion:part3}
For the OoIII group GCs, NGC 6388 and NGC 6441 \citep{2000ApJ...530L..41P,2009Ap&SS.320..261C}, which contain long-period and metal-rich RRab stars, a similar explanation can be applied as mentioned above. Their HBt values are very low, indicating a well-populated red HB star population. Nevertheless, both GCs show distinct blue HB extensions \citep{1997ApJ...484L..25R}, a feature that can efficiently generates variable stars via evolution, as observed in metal-poor GCs. From the data distribution presented in Figure \ref{Fig.7}, it is also evident that, in comparison to red HB stars, blue HB stars contribute significantly more to the formation of variable stars.

However, \citet{2002AJ....124..949P} conducted a study on variable stars in NGC 6388 and noted that the bluest HB stars evolve rapidly through the IS, spanning several $10^{5}$ years. This rapid evolution implies that such a short time frame is insufficient to account for the observed number of RRab stars. According to their research, the timescales of the targets in the IS should be less than 400,000 years (see the tracks labeled "c" in their Figure 11, less than 2 points in the IS). The BaSTI also provides the horizontal branch age of each evolutionary node, allowing us to evaluate the time experienced by these stars in the RRab IS. We estimated the timescales using the information provided by BaSTI. For $Z$ = 0.01, [Fe/H] = -0.60 (similar to the metallicity of NGC 6388), the timescale for the HB stars with 0.54 $M_{\odot}$ in the OoII region is 270,000 years, which is consistent with the results of \citet{2002AJ....124..949P}. However, it should be noted that in Panel (b) of Figure \ref{Fig.5}, the corresponding evolutionary track (red solid line) can be considered to be outside the OoII region. For the solid green line located in the center of the OoII region, the corresponding mass is 0.55 $M_{\odot}$, and the corresponding timescale is about 2.4 Myr. It can be seen that for relative metal-rich stars in their later stage of HB evolution, the evolutionary speed is very sensitive to the mass, and RRab stars located in the OoII region have more time to spend within the IS. Furthermore, it is important to note that \citet{2002AJ....124..949P} and \citet{2003AJ....126.1381P} have identified several BL Herculis stars (Type II Cepheids with periods of less than 4 or 5 days; \citealt{2017AcA....67..297S,2018AcA....68..315U}) in NGC 6388 and NGC 6441, which are generally believed to evolve away from the HB toward the AGB. Therefore, it is not surprising to find stars in the later stages of the HB evolutionary phase within these two GCs.

\begin{figure}
\centering
\includegraphics[width=0.45\textwidth]{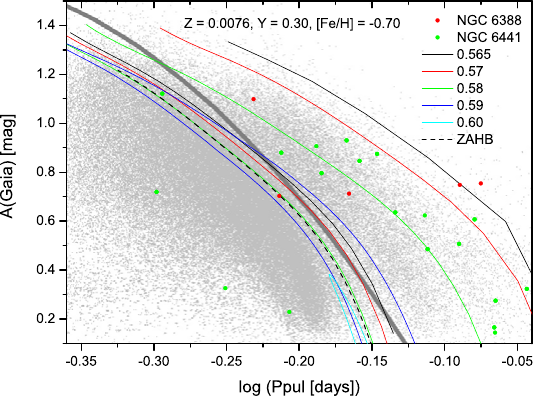}
\caption{Bailey diagram for Helium-enhanced model, i.e., Z = 0.0076, Y = 0.30, and [Fe/H] = -0.70. The gray background dots represent Gaia DR3 RRab stars, and the red and green solid dots represent RRab stars in NGC 6388 and NGC 6441, respectively. The black dashed line refers to the ZAHB. The solid lines represent the evolutionary tracks of stars with different masses. The thick gray solid lines present the division of OoI and OoII regions.} \label{Fig.8}
\end{figure}

\subsection{Helium-enchanced model}
The multiple populations and Helium-enhanced model have also been used to explain the Oosterhoff phenomenon in GCs. For instance, \citet{2014MNRAS.443L..15J} suggested that the RR Lyrae stars in OoIII GCs originate from more helium-rich third and later generations of stars. BaSTI provide ZAHB information for helium-enhanced scenarios. We plotted the Bailey diagram, which includes the ZAHB line and evolutionary tracks for a Helium-enhanced model (Y = 0.30 and [Fe/H] = -0.70\footnote{these parameters are similar to those considered in \citet{2014MNRAS.443L..15J}.}; see Figure \ref{Fig.8}). The results indicate that the ZAHB lines and evolutionary tracks are generally trend towards longer periods, predominantly located in the OoII region. However, when compared with the observed data (represented by red and green dots in Figure \ref{Fig.8}), it is evident that, even in the case of Helium-enhanced, most of the variables in the two OoIII GCs are in the later stages of HB.

In fact, the effects of evolution and Helium-enhanced are not mutually exclusive, and we can discuss which effect predominates in the Oosterhoff phenomenon. From the Bailey diagram of Gaia DR3 (gray background dots in Figure \ref{Fig.4}), it is apparent that most of the RRab stars in the Milky Way and neighboring galaxies are situated in the OoI region. This region can be effectively described by the canonical model, suggesting that the vast majority of HB stars are not helium-enhanced. Consequently, these stars must undergo evolution, which implies that there will be a certain proportion of HB stars will cross the IS from the blue side. These evolved stars are located in the OoII region on the Bailey diagram. Moreover, compared to the helium-enhanced model, the evolutionary effect can naturally account for the higher average rate of period changes observed in the OoII GCs \citep{1991ApJ...367..524L,2009Ap&SS.320..261C,2020A&A...635A..66P}. Of course, given the potential significance of helium in specific celestial structures such as GCs \citep{2016ApJ...827....2V,2019MNRAS.486.5895T}, or the Galactic bulge \citep{2018ApJ...853L..20M}, further detailed research is warranted.

\subsection{The ratio of OoI to OoII RRab stars}  \label{Sec:Discussion:part4}
As discussed earlier, we can use the HB age of each evolutionary node provided by BaSTI to evaluate the duration that these stars have spent in the RRab IS. Taking Panel (d) of Figure \ref{Fig.4} as an example ([Fe/H] = -1.62)\footnote{The reason for choosing this metal abundance is that it is closest to the mean metallicity of OoI stars, $\langle$[Fe/H]$\rangle$ = -1.59 \citep{2023MNRAS.525.5915Z}.}, the evolution of stars with a mass of 0.62 $M_{\odot}$ occurs in the OoII region, where they may remain in the RRab IS for approximately 5 Myr. In contrast, stars with a mass of 0.65 $M_{\odot}$ evolve in the OoI region, have spent 21 Myr in the early stage and an additional 10 Myr during the later redward stage. We make the following preliminary estimates: the average timescale for stars in the OoII region is 5 Myr, while the timescale for stars in the OoI region is approximately 30 Myr. Assuming that the number of observed stars is proportional to the aforementioned timescales, we expect the ratio of OoII to OoI stars to be around 1 : 6. According to the observational results from Gaia DR3, the actual ratio is 1 : 4.6 (31,156 : 144,194, using the thick gray solid line in Figure \ref{Fig.4} as the segmentation criterion). When considering samples from \citet{2020ApJS..247...68L} within this [Fe/H] range, the ratio is 1 : 6 (a total of 995 samples, with 852 and 143 samples belonging to OoI and OoII, respectively). From this perspective, the expected ratio is essentially consistent with the observed ratio.

\subsection{Future research}  \label{Sec:Discussion:part5}
In the present paper, we utilize empirical formulas to establish the relationship between the amplitude of RRab stars and their intrinsic $T_{\rm eff}$. The primary analysis focuses on this relationship. In fact, the correlation between pulsation amplitude and other parameters is also a topic worthy of further investigation. In this study, we simply assume that the amplitude is solely related to the $T_{\rm eff}$. However, some researchers suggest that additional factors may influence or correlate with amplitude. \citet{1990ApJ...350..155L} proposed that the amplitude is not exclusively dependent on temperature but also on [Fe/H]. The relationships documented in the in literature also indicate that the amplitude is correlated not only with the color index but also with [Fe/H] and/or pulsation period \citep{1992AJ....104..253C,2002AJ....124.1528P,2009AJ....138.1284K}. While amplitude is believed to be associated with the Oosterhoff type \citep{1999ApJ...515L..85C}, there exists an opposing view \citep{2014AJ....147...31M}. In recent years, researches on RR Lyrae stars utilizing spectral data from various survey projects have primarily concentrated on parameters such as elemental abundance and radial velocity (see series of papers of \citealt{2019ApJ...882..169F} and \citealt{2020ApJS..247...68L}). However, spectroscopic studies have yet to systematically analyze other physical parameters (e.g., $T_{\rm eff}$, $\log$ g). Scholars may be able to directly obtain these parameters during different pulsation phases through spectroscopy \citep{2014MNRAS.445.4094F,2017ApJ...835..187C}, subsequently calaculating their average values. These averages can then be compared with physical and pulsation parameters to identify specific statistical patterns and correlations. Additionally, similar analyses can be conducted using the photometric color indices \citep{2017ApJ...834..160N}; however, addressing the influence of interstellar extinction presents a complex challenge.

\section{Summary}  \label{Sec:Conclusion}

The mechanism underlying the Oosterhoff dichotomy represents a significant open question in the field of RR Lyrae star research, with evolutionary effects frequently cited as a potential explanation. By examining the relationships between the pulsation parameters of these stars and their evolutionary characteristics, we construct an "HR diagram" using pulsation parameters as variables \citep{1997A&AS..121..327B,2020ApJ...896L..15B}. We hypothesized that the pulsation amplitude is a function of the average $T_{\rm eff}$ and derived their relationship from empirical formulas \citep{2014AJ....147...31M,2016A&A...595A.133C}. Utilizing these relationships, we converted the evolutionary parameters provided by BaSTI into corresponding pulsation parameters and illustrated them on the Bailey diagram. Ultimately, we obtained the following results.

1. The distributions of ZAHB lines indicate that all pulsating ZAHB stars belong to the OoI type, with their amplitudes decreasing as the $P_{\rm pul}$ increase (see Figure \ref{Fig.2}). When [Fe/H] $<$ -1, there is minimal variation in the distribution of ZAHB lines for different [Fe/H] values ($\Delta \log P_{\rm pul}/\Delta$[Fe/H] = -0.003). However, when [Fe/H] $>$ -1, the differences in luminosity become significant, resulting in the period shift phenomenon observed in the Bailey diagram ($\Delta \log P_{\rm pul}/\Delta$[Fe/H] = -0.087, see Figure \ref{Fig.3}).

2. The evolutionary tracks indicate that the vast majority of OoI stars are HB stars in the early stages of HB evolution, while all OoII stars are in the later stages (see Figures \ref{Fig.4} and \ref{Fig.5}). This confirms the previous evolutionary viewpoints (see Section \ref{sec:intro}). The mass of RRab stars decreases with increasing metallicity (see Figure \ref{Fig.6}), which aligns with the general understanding \citep{1998A&A...333..571J}. However, for the same metallicity, the mass distribution should exhibit a certain degree of broadening, and the mass of OoII stars is less than that of OoI stars (see Figure \ref{Fig.6}). The $T_{\rm eff}$ of OoII stars are higher than the blue boundary of the IS at their ZAHB, only entering the IS at higher luminosities in the later stages (see Figure \ref{Fig.7}). These results suggest that evolutionary effects (mass, metallicity, luminosity, $T_{\rm eff}$, and HB age) are the direct cause of the Oosterhoff period dichotomy.

3. Among these physical parameters of evolution, metallicity plays an important role in another aspect. It regulates the proportion of evolutionary pathways by influencing HB morphology of GCs, which results in the statistical Oosterhoff phenomenon (see Section \ref{Sec:Discussion:part2}). Moreover, regardless of whether there is helium enhancement, the long-period, metal-rich RRab stars in the two OoIII GCs, NGC 6388 and NGC 6441, are in the late stages of HB evolution (see Panel (b) of Figure \ref{Fig.5} and Figure \ref{Fig.8}). From this perspective, these two GCs can be classified as belonging to the OoII group.

Our study corroborates previous finding from a novel perspective, demonstrating that the direct distinction between OoI and OoII stars lies in their evolutionary stages. Furthermore, we establish the significant role of metallicity: it acts as the first parameter governing HB morphology \citep{2024MNRAS.534.3654P}, which subsequently determines the evolutionary stage at which HB stars predominantly enter the IS. The observed statistical correlation between the pulsation periods and metallicity \citep{2021ApJ...919..118F}, can also be attributed to the influence of these factors. Research on the Oosterhoff phenomenon will continue unabated, as it may provide crucial insights or breakthroughs in the exploration of the Milky Way and nearby dwarf galaxies. However, before progressing further, it is essential to renew our focus on the evolutionary effects that directly contribute to the Oosterhoff phenomenon and to clarify the significant role of metallicity within this context.

\section{acknowledgments}

We sincerely thank the reviewers for their constructive comments and suggestions, which have significantly improved the quality of this work. This work is supported by the Yunnan Fundamental Research Projects (grant Nos. 202503AP140013, 202501AS070055, 202401AS070046, 202301AT070352, 202201AT070187), the International Partnership Program of Chinese Academy of Sciences (No. 020GJHZ2023030GC), the China Manned Space Program with grant no. CMS-CSST-2025-A16, the CAS "Light of West China" Program and "Yunnan Revitalization Talent Support Program" in Yunnan Province.

This research uses data from \citet{2001AJ....122.2587C}, \citet{2006ApJ...642..797P}, \citet{2017yCat.5150....0C}, \citet{2019ApJS..245...34X}, \citet{2020ApJS..247...68L}, \citet{2022yCat.1358....0G} and \citet{2023A&A...674A..18C}. We express our gratitude for their excellent work. Without their contributions, this paper would not have been possible to complete. This research made use of the cross-match service provided by CDS, Strasbourg. This work has made use of data from the European Space Agency (ESA) mission {\it Gaia} (\url{https://www.cosmos.esa.int/gaia}), processed by the {\it Gaia} Data Processing and Analysis Consortium (DPAC, \url{https://www.cosmos.esa.int/web/gaia/dpac/consortium}). Funding for the DPAC has been provided by national institutions, in particular the institutions participating in the {\it Gaia} Multilateral Agreement.

\section{Date availability statements}

The data used in this article are from BaSTI, and the original data can be obtained through the following links query:
\url{http://albione.oa-teramo.inaf.it/}



\bibliographystyle{mnras}
\bibliography{Ref} 









\bsp	
\label{lastpage}
\end{document}